\documentclass[superscriptaddress,aps,pra,twocolumn,showpacs,nofootinbib,longbibliography]{revtex4-2}
\usepackage{amsmath,amssymb,amsthm}
\usepackage{easybmat}
\usepackage[colorlinks=true,citecolor=blue,urlcolor=blue]{hyperref}
\usepackage[pdftex]{graphicx}
\usepackage{times,txfonts}
\usepackage{braket}
\usepackage{color}
\usepackage{natbib}
\newcommand{\be}{\begin{equation}}
	\newcommand{\ee}{\end{equation}}
\newcommand{\ba}{\begin{eqnarray}}
	\newcommand{\ea}{\end{eqnarray}}
\newcommand{\ketbra}[2]{|#1\rangle \langle #2|}

\begin{document}
	
	\title{A study of the quasi-probability distributions of the Tavis-Cummings model under different quantum channels}

	\author{Devvrat Tiwari\textsuperscript{}}
	\email{devvrat.1@iitj.ac.in}
	\author{Subhashish Banerjee\textsuperscript{}}
	\email{subhashish@iitj.ac.in }
	\affiliation{Indian Institute of Technology Jodhpur-342030, India\textsuperscript{}}
	


\date{\today}

\begin{abstract}
	
	We study the dynamics of the spin and cavity field of the Tavis-Cummings model using quasi-probability distribution functions and the second-order coherence function, respectively. The effects of (non)-Markovian noise are considered. The relationship between the evolution of the cavity photon number and spin excitation under different quantum channels is observed. The equal-time second-order coherence function is used to study the sub-Poissonian behavior of light and is compared with the two-time second-order coherence function in order to highlight the (anti)-bunching properties of the cavity radiation.   
\end{abstract}

\maketitle


\section{\label{sec:Intro}Introduction }  
Interactions between atoms and the cavity field play a critical role in science and technology and provide a foundation to quantum optics~\cite{loudon, mandel-wolf, scully-zubairy, gs-agarwal}. They are at the core of numerous developments in spectroscopy, quantum information processing, sensing, and lasers, among others. The first exactly solvable quantum mechanical model of a single two-state atom interacting with a cavity mode of an electromagnetic field was given by Jaynes and Cummings~\cite{jaynes-cummings, shore-knight}. It was developed to examine the processes of spontaneous emission and absorption of photons in a cavity as well as to detect the presence of Rabi oscillations in atomic excitations. Experimental verification of the Jaynes-Cummings model~\cite{meschede-walther, haroche-raimond} significantly increased its importance. The Tavis-Cummings (TC) model~\cite{tavis-cummings, tavis-cummings2, kurucz-molmer, bogoliubov, klimov-chumakov, scully-zubairy, hsdhar}, a multi-atom generalization of the Jaynes-Cummings model, is of fundamental importance in the quest to understand atom-field interactions. It appears with different variations in quantum physics and has a resemblance with the Dicke model~\cite{dicke, garraway} under dipole and rotating wave (RW) approximations, modulo the different coupling strengths between atoms and the cavity field and the inhomogeneous transition frequencies of the individual atoms.

Cavity quantum electrodynamics (cavity-QED) studies the properties of atoms interacting with photons in cavities~\cite{purcell-qed, berman-qed, haroche-qed, walther-varcoe-qed}. Many theoretical models can now be realized in laboratories thanks to advances in cavity-QED experiments during the previous few decades. It is now possible to control the isolated evolution of a few atoms coupled to a single mode inside a cavity. The TC model has been realized in a number of experiment~\cite{tc-wang, tc-sillanpaa, majer-chow, tc-fink}, and their applications can be found in~\cite{majer-chow, tc-wang, tc-appl-woerkom, tc-astner, tc-deng} and references therein.

The notion of phase space is a very useful concept in analyzing the dynamics of classical systems. However, a straightforward extension to the phase space domain in quantum mechanics is hampered because of the uncertainty principle. In spite of this, quasi-probability distribution functions (QDs) for quantum mechanical systems analogous to their classical counterparts can be constructed~\cite{klauder-sudarshan, scully-zubairy, schleich, thapliyal-banerjee, stratonovich, qd-thapliyal, qd-gs-agarwal, rr-puri, klimov-chumakov}. These QDs are extremely helpful because they offer a quantum-classical relationship and make it easier to calculate quantum mechanical averages that are analogous to classical phase space averages. However, the QDs are not probability distributions since they may also have negative values. The first such QD developed was the Wigner $W$ function~\cite{w-wigner, w-moyal, w-hillery, w-kim, w-miranowicz, w-opatrny}. The $P$ function is a different, well-known QD that acts as a witness to quantumness in the system. It can become singular for some quantum states. The $W$ and $P$ functions, along with the $Q$ function~\cite{q-mehta-sudarshan, q-kano, q-husimi}, are used in the present study. The problem of operator orderings is closely tied to these QDs. As a result, the $P$ and $Q$ functions are related to normal and anti-normal orderings, respectively, while the $W$ function is connected to symmetric operator ordering~\cite{Louisell}. 

The tremendous interest in these QDs can be attributed to a number of factors. We can use them to identify a state's non-classical characteristics (quantumness in the system)~\cite{ryu-lim}. Values of $P$ function that are non-positive precisely characterize a non-classical state. $P$ function's non-positivity is a necessary and sufficient condition for non-classicality in a system, although other QDs provide only sufficient criteria.
Tasks that are impossible in a classical state can be accomplished using a non-classical state. Numerous investigations on non-classical states, such as those on squeezed, antibunched, and entangled states, were motivated by this~\cite{pathak-elements}. Interestingly, many of these applications have been developed using spin-qubit systems.


A realistic quantum system is subjected to the influence of the environment. These interactions significantly change the system's dynamics and result in the loss of information from the system to the environment. The theory of open quantum systems (OQS)~\cite{bruer-petrrucione, sbbook, weiss} provides a framework to study the impact of the environment on a quantum system. Open quantum system ideas cater to a broad spectrum of disciplines~\cite{Louisell, caldeira-leggett1983, GrabertSchrammIngold, sbqbm, sbsterngerlach, sbrichard, sbjavidprd, sbkhushboocoherence, sbunruh1, sbunruh2, reactioncoordinaterefs1, reactioncoordinaterefs2, reactioncoordinaterefs3, blhu, sgad, plenio}. In many circumstances, the dynamics of an OQS may be characterized using a Markov approximation, which assumes that the environment instantaneously recovers from its contact with the system, resulting in a continuous flow of information from the system to the environment. However, growing technical as well as technological advances are pushing the study into regimes beyond Markovian approximation. A neat separation between system and environment time scales can no longer be expected in many of these circumstances, resulting in non-Markovian behavior~\cite{breuer1,laine,rivas,rivas1,vasile, devega-alonso,lu,luo,fanchini,titas,haseli,hall1,hall2, sam1-banerjee, kumar2018enhanced, kumar2018non, nmad-channel, phase-covariant, semi-markov-channel}.


With the motivation to understand the impact of noise, both Markovian and non-Markovian, in the context of cavity-QED, specifically on the Tavis-Cummings model, we will make use of the $W$, $P$, and $Q$ QDs to study the dynamics of the spin system. For the characterization of the cavity field, the second-order coherence function~\cite{loudon,mandel-wolf} will be used.

This paper is organized as follows: in Sec.~\ref{sec:Model}, we present, briefly, the Tavis-Cummings model and a master equation to understand the open system dynamics of the TC model. A brief discussion on QDs is provided in Sec.~\ref{sec:QDs} followed by a study of the dynamics of the spin system using QDs in Sec.~\ref{sec:spin-dynamics} and the dynamics of the cavity field in Sec.~\ref{sec:cavity-dynamics}, under ambient noisy conditions. We then make our conclusions.

\section {\label{sec:Model}The Model}
Here we consider the Tavis-Cummings model with $N$ two-level atoms or spins with inhomogeneous transition frequencies coupled to a single mode field cavity with different coupling strengths. Under the dipole and rotating wave approximations (RWA), we can write the Hamiltonian of the system (with $\hbar=1$) as
\begin{equation}
    H = \frac{1}{2}\sum_{k=1}^N\omega_k\sigma^z_k + \omega_c a^{\dagger}a + \sum_{k=1}^N g_k(\sigma_k^+a + \sigma_k^-a^{\dagger}),
    \label{Eq:Hamiltonian}
\end{equation}
where $\omega_c$ is the resonance frequency of the cavity field, and $\omega_k$ and $g_k$ are the transition frequencies and coupling strength for the $k^{th}$ spin. $\sigma^z_k$, $\sigma^+_k$ and $\sigma^-_k$ are the Pauli spin-$1/2$ operators. 

To account for the losses in the spin and cavity system due to various dissipative processes, such as spontaneous emission and cavity decay due to imperfections in the cavity, we use the tools of open quantum systems. To this end, the losses can be modeled by the following master equation
\begin{align}
    \frac{d\rho(t)}{dt} &= -i[H, \rho(t)] + \frac{1}{2}\sum_{k=1}^N \bigg[\gamma_k (N_{th, k}+1) (2\sigma_k^-\rho(t)\sigma_k^+ \nonumber \\
    &- \{\sigma_k^+\sigma_k^-, \rho(t)\}) + \gamma_k N_{th, k}(2\sigma_k^+\rho(t)\sigma_k^- - \{\sigma_k^-\sigma_k^+, \rho(t)\})\bigg]\nonumber \\
    &+\frac{1}{2} \left[ \kappa (N_{th, k}+1) (2a \rho(t) a^{\dagger} - \{a^{\dagger} a, \rho(t)\})\right.\nonumber \\
    &+ \left.\kappa N_{th, k} (2a^{\dagger} \rho(t) a - \{a a^{\dagger}, \rho(t)\})\right].
    \label{Eq:GKSL-master-eq}
\end{align}
Here $\kappa$ corresponds to the cavity decay rate, $\gamma_k$ corresponds to the spontaneous emission rate, and $N_{th}$ is the average number of thermal photons. Constant values of these decay rates generate a semi-group type of Lindblad equation, called the GKSL master equation~\cite{gksl, Lindblad}, whereas their time dependence typically models non-Markovian scenarios. In the subsequent sections, we will see the impact of various types of noise (Markovian and non-Markovian) on the dynamics of the system. 

\section{\label{sec:QDs}Quasi-probability Distribution Functions}
Here we briefly discuss the Wigner $W$, $P$, and $Q$ quasi-probability distribution functions (QDs) to be used subsequently.  

\subsection{\label{sec:W-func}The Wigner $W$ function}
The Wigner $W$ function for a single spin-$j$ state, as a function of polar and azimuthal angles expanded over special harmonics, can be given as
\begin{equation}
    W(\theta, \phi) = \bigg(\frac{2j+1}{4\pi}\bigg)\sum_{\mu, \eta}\rho_{\mu \eta}Y_{\mu \eta}(\theta, \phi),
    \label{W-func-1qubit}
\end{equation}
where $\mu = 0, 1, \dots,2j$ and $\eta = -\mu, -\mu+1,\dots,0,\dots,\mu-1,\mu$, and 
\begin{equation}
    \rho_{\mu \eta} = \mathrm{Tr}\big[T_{\mu \eta}^{\dagger}\rho\big].
    \label{Eq:rkq}
\end{equation}
Further, $Y_{\mu \eta}$ are the spherical harmonics and $T_{\mu \eta}$ are the multipole operators~\cite{varshalovich, zare} given by
\begin{equation}
    T_{\mu \eta} = \sum_{m, m'}(-1)^{j-m}(2\mu +1)^{1/2}\begin{pmatrix}
    j&\mu &j\\
    -m&\eta&m'
    \end{pmatrix}\ket{j, m}\bra{j, m'},
    \label{Eq:tkq}
\end{equation}
where $\begin{pmatrix}j_1&j_2&j\\m_1&m_2&m\end{pmatrix} = \frac{(-1)^{j_1-j_2-m}}{\sqrt{2j+1}}\bra{j_1m_1j_2m_2}{j-m}\rangle$ is the Wigner $3j$ symbol and $\bra{j_1m_1j_2m_2}{j-m}\rangle$ is the Clebsch-Gordon coefficient. The multipole operators are orthogonal to each other, and they form a complete set with $T_{\mu \eta}^{\dagger} = (-1)^\eta T_{\mu, -\eta}$. The $W$ function satisfies the normalization condition
\begin{equation}
    \int W(\theta, \phi) \sin \theta d\theta d \phi = 1,
    \label{Eq:w-normal}
\end{equation}
and $W^* (\theta, \phi) = W(\theta, \phi)$. In a similar way, we can write the $W$ function for an $N$ particle system, each with spin-$j$ as
\begin{align}
    W(\theta_1, \phi_1,\dots, \theta_N, \phi_N) &= \bigg(\frac{2j+1}{4\pi}\bigg)\sum_{\mu _1, \eta_1}\sum_{\mu_2, \eta_2}...\sum_{\mu_N, \eta_N}\nonumber \\
    &\rho_{\mu_1\eta_1\mu_2\eta_2,\dots,\mu_N\eta_N}Y_{\mu_1\eta_1}(\theta_1, \phi_1)\nonumber \\
    &\times Y_{\mu_2, \eta_2}(\theta_2, \phi_2)\dots Y_{\mu_N\eta_N} (\theta_N, \phi_N),
    \label{Eq:W-func-Nqubit}
\end{align}
where $\rho_{\mu_1\eta_1\mu_2\eta_2\dots \mu_N\eta_N} = \mathrm{Tr}\big\{\rho T^{\dagger}_{\mu_1\eta_1}T^{\dagger}_{\mu_2\eta_2}\dots T^{\dagger}_{\mu_N\eta_N}\big\}$, satisfying the normalization condition 
\begin{align}
    \int W(\theta_1, \phi_1,\dots, \theta_N, \phi_N) \sin\theta_1 \sin\theta_2\dots \sin\theta_N d\theta_1 d\phi_1 \nonumber \\ \times d\theta_2 d\phi_2  \dots d\theta_N d\phi_N = 1.
    \label{Eq:w-normal-Nqubit}
\end{align}
\subsection{\label{sec:P-func}The $P$ function}
The $P$ function for a single spin-$j$ particle is defined as
\begin{equation}
    \rho = \int d\theta d\phi P(\theta, \phi)\ketbra{\theta, \phi}{\theta, \phi},
\end{equation}
and can be shown to be 
\begin{align}
    P(\theta, \phi) &= \sum_{\mu, \eta}\rho_{\mu \eta}Y_{\mu \eta} (\theta, \phi)\left(\frac{1}{4\pi}\right)^{1/2} (-1)^{\mu-\eta}\nonumber \\
    &\times\Bigg(\frac{(2j-\mu)!(2j+\mu+1)!}{(2j)!(2j)!}\Bigg)^{1/2}.
    \label{Eq:P-func-1qubit}
\end{align}
Here $\ket{\theta, \phi}$ is the atomic coherent state~\cite{arecchi} which in terms of Wigner-Dicke states $\ket{j,m}$ can be expressed as
\begin{align}
    \ket{\theta, \phi} = \sum_{m=-j}^j \begin{pmatrix}2j\\m+j\end{pmatrix}^{1/2}\sin^{j+m}\left(\frac{\theta}{2}\right)\cos^{j-m}\left(\frac{\theta}{2}\right)e^{-i(j+m)\phi}\ket{j,m}. 
\end{align}
Moreover, the $P$ function for $N$ spin-$j$ particles is 
\begin{widetext}
\begin{align}
P(\theta_1, \phi_1,\dots\theta_N, \phi_N) &= \sum_{\mu_1, \eta_1}\sum_{\mu_2, \eta_2}...\sum_{\mu_N, \eta_N}\rho_{\mu_1\eta_1\mu_2\eta_2,\dots,\mu_N\eta_N}Y_{\mu_1\eta_1}(\theta_1, \phi_1) Y_{\mu_2, \eta_2}(\theta_2, \phi_2)\dots Y_{\mu_N\eta_N} (\theta_N, \phi_N) (-1)^{\mu_1 - \eta_1 +\mu_2 - \eta_2 + \dots \mu_N - \eta_N}\nonumber \\ 
&\times \left(\frac{1}{4\pi}\right)^{N/2}\left(\frac{\sqrt{(2j-\mu_1)!(2j-\mu_2)!\dots(2j-\mu_N)!(2j+\mu_1+1)!(2j+\mu_2+1)!\dots(2j+\mu_N+1)!}}{[(2j)!(2j)!]^{N/2}}\right).
\label{Eq:P-func-Nqubit}
\end{align}
\end{widetext}

\subsection{\label{sec:Q-func} The $Q$ function}
The $Q$ function for a single spin-$j$ state is defined as
\begin{equation}
    Q(\theta, \phi) = \frac{2j+1}{4\pi}\bra{\theta, \phi}\rho\ket{\theta, \phi},
\end{equation}
and can be shown to be
\begin{align}
    Q(\theta, \phi) &=\bigg(\frac{1}{4\pi}\bigg)^{1/2}\sum_{\mu, \eta}\rho_{\mu \eta}Y_{\mu \eta} (\theta, \phi) (-1)^{\mu-\eta}(2j+1) \nonumber \\
    &\times \bigg(\frac{(2j)!(2j)!}{(2j-\mu)!(2j+\mu+1)!}\bigg)^{1/2}.
    \label{Eq:Q-func-1qubit}
\end{align}
For $N$ spin-$j$ particles, the normalized $Q$ function is
\begin{widetext}
\begin{align}
    Q(\theta_1, \phi_1,\dots\theta_N, \phi_N) &= \sum_{\mu_1, \eta_1}\sum_{\mu_2, \eta_2}...\sum_{\mu_N, \eta_N}\rho_{\mu_1\eta_1\mu_2\eta_2,\dots,\mu_N\eta_N}Y_{\mu_1\eta_1}(\theta_1, \phi_1) Y_{\mu_2, \eta_2}(\theta_2, \phi_2)\dots Y_{\mu_N\eta_N} (\theta_N, \phi_N) (-1)^{\mu_1 - \eta_1 +\mu_2 - \eta_2 + \dots \mu_N - \eta_N} \nonumber \\ 
    &\times \left(\frac{(2j+1)^N}{(4\pi)^{N/2}}\right) \left(\frac{[(2j)!(2j)!]^{N/2}}{\sqrt{(2j-\mu_1)!(2j-\mu_2)!\dots(2j-\mu_N)!(2j+\mu_1+1)!(2j+\mu_2+1)!\dots(2j+\mu_N+1)!}}\right).
    \label{Eq:Q-func-Nqubit}
\end{align}
\end{widetext}

\section{\label{sec:spin-dynamics}Dynamics of the spin system: Quasi-probability Distributions}
We consider the dynamics of the spin system inside the cavity under the influence of various quantum channels and study their $W$, $P$, and $Q$ quasi-probability distributions. In order to bring out the impact of non-Markovian effects on the evolution of the spin, we make a comparison with the corresponding analysis under the influence of the GKSL master equation. We consider $N=4$ and take the initial state of the spin-cavity system to be $\rho = \rho_a(0) \otimes \ketbra{\alpha}{\alpha}$, where $\rho_a(0) = \ketbra{g}{g}$ ($\ket{g}$ denotes the ground state of the atomic system), and $\ket{\alpha}$ ($\alpha = |\alpha|e^{i\zeta}$) is the field coherent state, such that $\ket{\alpha} = e^{-\braket{n}/2}\sum_{n=0}^\infty \frac{(\sqrt{\braket{n}})^n}{\sqrt{n!}}e^{in\zeta}\ket{n}$, where $n$ is the photon number in the Fock state $\ket{n}$. In the numerics below, we choose $\braket{n} = 6$ and $\zeta = \pi/2$. 

\subsection{\label{sec:noisy-channels}Impact of noisy channels}

We now study the impact of a number of noisy channels, both Markovian as well as non-Markovian, on the dynamics of the spin system inside the cavity.

\subsubsection{\label{sec:SGAD}Squeezed Generalized Amplitude Damping Channel}
We consider the dynamics of the spin system impacted by the squeezed generalized amplitude damping (SGAD) channel~\cite{sgad, Omkar-sgad}, which is a GKSL type of semi-group evolution (Markovian in nature). It is worth noting that here and in all the master equations treated below, the cavity loss is accounted for by the term corresponding to $\kappa$. The master equation for the evolution of the spin-cavity system is
\begin{align}
    \frac{d\rho(t)}{dt} &=  -i[H, \rho(t)] + \frac{1}{2}\sum_{k=1}^N \sum_{j=1}^2 (2R_{jk}\rho(t) R^{\dagger}_{jk} - R_{jk}^{\dagger}R_{jk}\rho(t) \nonumber \\
    &- \rho(t) R_{jk}^{\dagger}R_{jk})+\frac{1}{2} \kappa (2a \rho(t) a^{\dagger} - a^{\dagger} a\rho(t) - \rho(t) a^{\dagger}a), 
    \label{Eq:SGAD-masterEq}
\end{align}
where $R_{1k}=(\gamma_k(N_{th,k}+1))^{1/2}R$ and $R_{2k} = (\gamma_kN_{th,k})^{1/2}R$ with $R = \sigma_k^- \cosh(r)+e^{i\Phi}\sigma_k^+ \sinh(r)$ and $N_{th,k} = \frac{1}{e^{h\omega_k/k_BT} - 1}$. Also, $r$ and $\Phi$ are the bath squeezing parameters. 
\begin{figure}[h]
	\includegraphics[height=75mm,width=1\columnwidth]{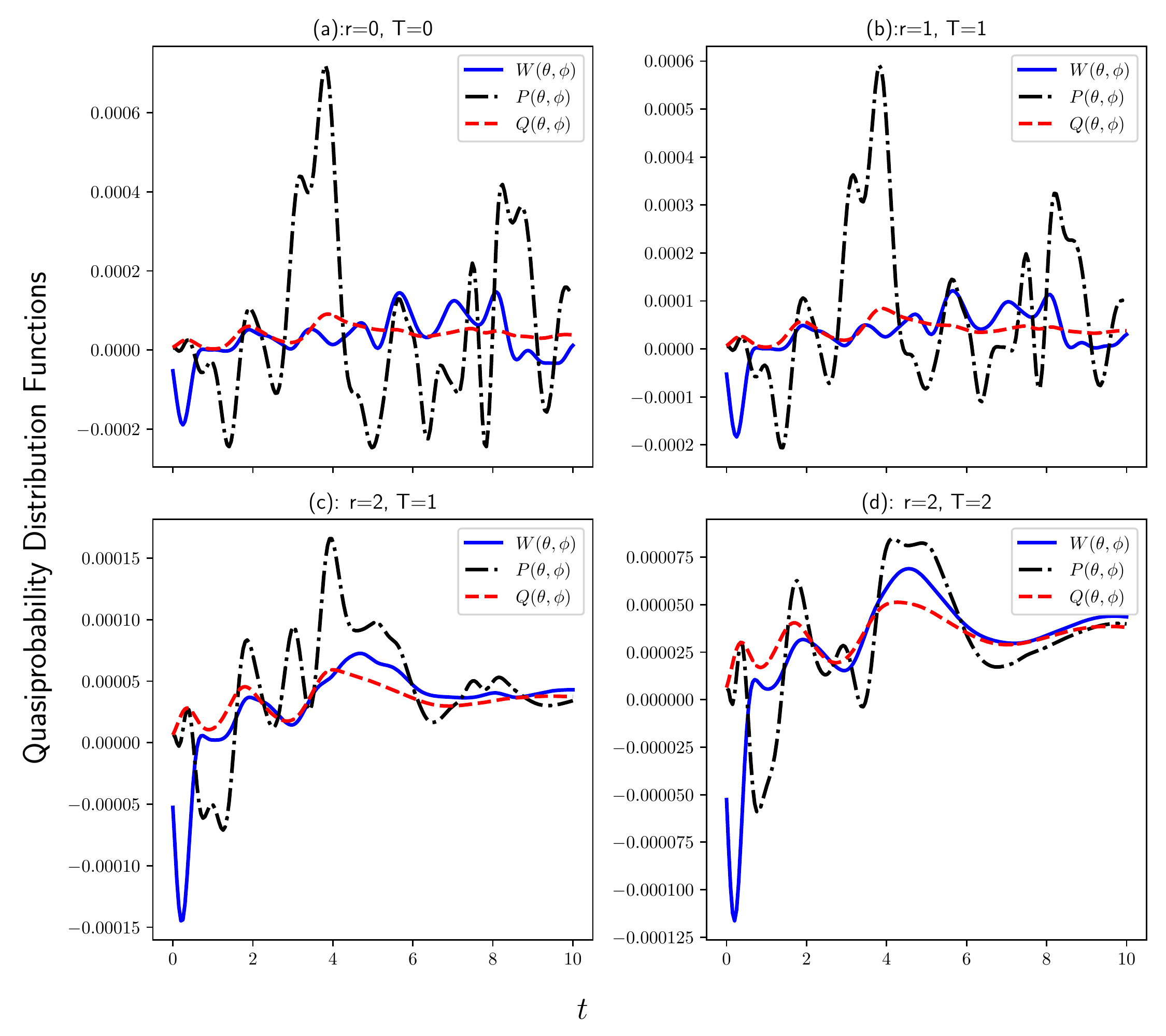}
	\caption{Variation of the $W$, $P$, and $Q$ functions for the spin system with time under the squeezed generalized amplitude damping channel. Here we have chosen $\omega_1 = \omega_4 = 1.11$, $\omega_2 = \omega_3 = 1.15$,  $\omega_c = 1.15$, $\Phi=\pi/4$, $g_1 = g_4 = 0.55$, $g_2 = 0.52$, $g_3 = 0.5$, $\gamma_k = 0.01g_k$ and $\kappa = 0.01$. The subplots (a), (b), (c), and (d) have different values of the squeezing parameter $r$ and temperature $T$. Also, $\theta_i = \{\pi/4, 3\pi/5, 2\pi/3, 3\pi/4\}$ and $\phi_i = \{3\pi/4, \pi/3, \pi/4, \pi/6\}$.}
	\label{fig:tc_simp_vs_squeezed}
\end{figure}
After the evolution of the system $\rho$ through the SGAD channel, we trace out the cavity degrees of freedom and use Eqs. (\ref{Eq:W-func-Nqubit}), (\ref{Eq:P-func-Nqubit}) and (\ref{Eq:Q-func-Nqubit}) on the spin system to calculate their $W$, $P$ and $Q$ functions, respectively. The QDs for the case of temperature $T$ and squeezing parameter $r$ equal to zero are plotted in Fig. \ref{fig:tc_simp_vs_squeezed}(a). It is observed that the $P$ and $W$ functions frequently take negative values, indicating quantumness in the system. Further, with an increase in the value of $r$ and $T$ in Figs. \ref{fig:tc_simp_vs_squeezed}(b), \ref{fig:tc_simp_vs_squeezed}(c), \ref{fig:tc_simp_vs_squeezed}(d), it is observed that the negativity of QDs decreases. This indicates depletion of non-classicality in the system with an increase in the squeezing parameter $r$ or temperature $T$. 

\begin{figure}[h]
	\includegraphics[width=1\columnwidth]{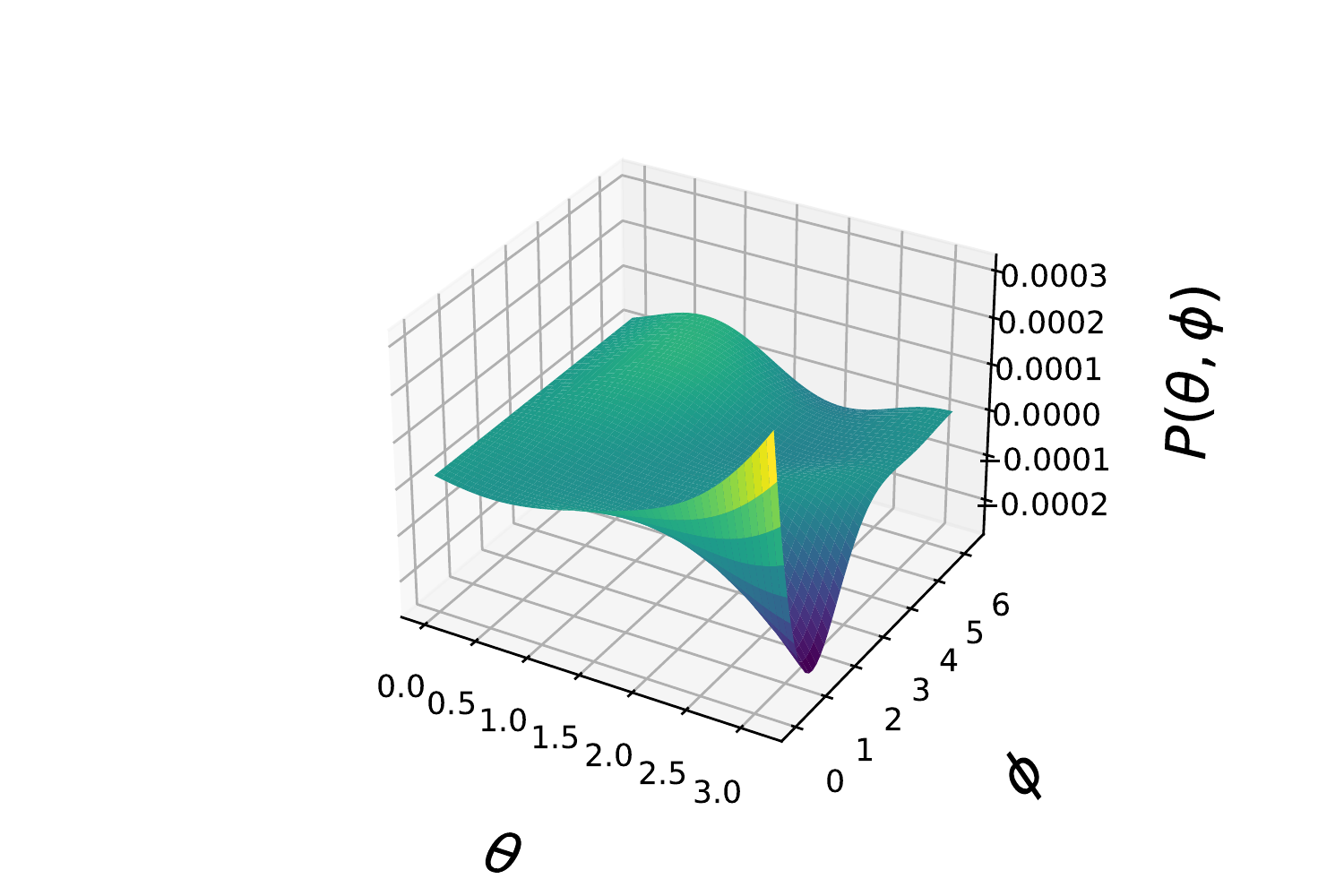}
	\caption{Variation of the $P$ function for the spin system with parameters $\theta$ and $\phi$ for a given time under the squeezed generalized amplitude damping channel. The parameters have the following values: $\omega_1 = \omega_4 = 1.11$, $\omega_2 = \omega_3 = 1.15$, $\omega_c = 1.15$, $g_1 = g_4 = 0.55$, $g_2 = 0.52$, $g_3 = 0.5$, $\gamma_i = 0.01g_i$, $r = 0$ and $T=0$, and $\kappa  =0.01$. }
	\label{fig:P_vary_theta_phi}
\end{figure}

It is worth mentioning that the negativity of $W$ and $P$ functions also depend on the choice of the system parameters $\theta_i$ and $\phi_i$ for each atom. Here we have run simulations on many values of $\theta_i$ and $\phi_i$ and have chosen the values of $\theta_i$ and $\phi_i$ showing greater negativity. Given that there are two parameters ($\theta_i$ and $\phi_i$) for each atom, therefore, for the case considered here ($N = 4$), we have a set of 8 parameters. It is not possible to show the variation of $W$ or $P$ function for all the 8 parameters here. However, to illustrate the dependence of negativity of $W$ or $P$ function over the parameters $\theta_i$ and $\phi_i$, we have chosen $\theta_1 = \frac{\theta}{4}, \theta_2 = \frac{3\theta}{5}, \theta_3 = \frac{2\theta}{3}, \theta_4 = \frac{3\theta}{4}$ and $\phi_1 = \frac{3\phi}{4}, \phi_2 = \frac{\phi}{3}, \phi_3 = \frac{\phi}{4}, \phi_4 = \frac{\phi}{6}$ and evolved the $P$ function with $\theta$ and $\phi$ for a given time. This is depicted in Fig. \ref{fig:P_vary_theta_phi}, where we can see the negativity of the $P$ function varies as we change $\theta$ and $\phi$.

Next, we consider the impact of non-Markovian noise on the spin system inside the cavity. To this effect, we consider the following non-Markovian channels.  

\subsubsection{\label{sec:phase-eternal-NM} Phase covariant eternal non-Markovian channel}
We now discuss the phase covariant eternal CP-indivisible dynamics~\cite{phase-covariant} of the spin system using the master equation
\begin{align}
    \frac{d\rho(t)}{dt} &= -i[H,\rho(t)] + \sum_{k=1}^N\left[\gamma_{1k}(t) \mathcal{L}_{1k}(\rho(t)) + \gamma_{2k}(t)\mathcal{L}_{2k}(\rho(t))\right. \nonumber \\
    & + \left.\gamma_{3k}(t)\mathcal{L}_{3k}(\rho(t))\right]+\frac{1}{2} \kappa (2a \rho(t) a^{\dagger} - a^{\dagger} a\rho(t) - \rho(t) a^{\dagger}a),
    \label{Eq:phase-cov-eternal}
\end{align}
where
\begin{equation}
\begin{aligned}
    \mathcal{L}_{1k} &= \sigma_k^+\rho(t)\sigma_k^- - \frac{1}{2}\{\rho(t), \sigma_k^-\sigma_k^+\}, \\
    \mathcal{L}_{2k} &= \sigma_k^-\rho(t)\sigma_k^+ - \frac{1}{2}\{\rho(t), \sigma_k^+\sigma_k^-\}, \\
    \mathcal{L}_{3k} &= \sigma_k^z\rho(t)\sigma_k^z - \rho(t). 
    \label{Eq:L1-L2-L3}
\end{aligned}
\end{equation}
Further, 
\begin{equation}
\begin{aligned}
    \label{phase-covariant-gammas}
    \gamma_{1k}(t)&= \nu_k(1+q), \\
    \gamma_{2k}(t)&= \nu_k(1-q), \\
    \gamma_{3k}(t)&= - \frac{\nu_k(1 - q^2)\sinh(2\nu_k t)}{2[1+q^2+(1-q^2)\cosh(2\nu_k t)]}, 
\end{aligned}
\end{equation}
such that $\nu_k>0$ and $|q|<1$. Here $\gamma_{1k}$, $\gamma_{2k}$, and $\gamma_{3k}$ denote the energy gain, energy loss, and pure dephasing rates, respectively. The cavity degrees of freedom are now traced out from the solution of Eq. (\ref{Eq:phase-cov-eternal}) and subsequently used to calculate the QDs of the spin system using Eqs. (\ref{Eq:W-func-Nqubit}), (\ref{Eq:P-func-Nqubit}) and (\ref{Eq:Q-func-Nqubit}). 
\begin{figure}[h]
	\includegraphics[height=60mm,width=1\columnwidth]{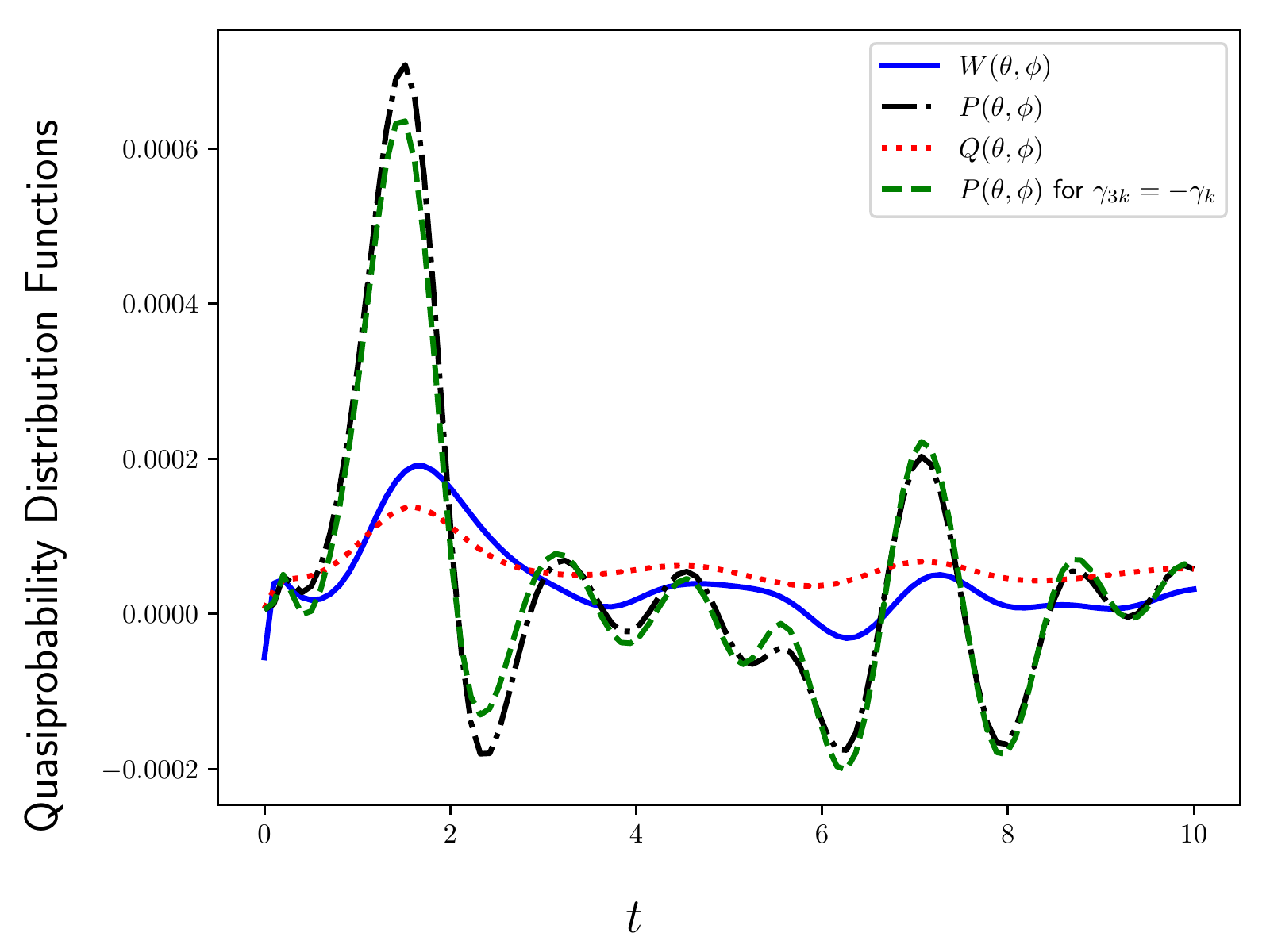}
	\caption{Variation of the $W$, $P$, and $Q$ function for the spin system with time after evolution through phase covariant eternal non-Markovian channel given in Eq. (\ref{Eq:phase-cov-eternal}). The green dashed curve denotes the variation of the $P$ function when the pure dephasing rate in Eq. (\ref{phase-covariant-gammas}) $\gamma_{3k}(t)$ is chosen to be in the limiting range of $t \to \infty$, which becomes $-\nu_k/2$. We have chosen the parameters to be $\omega_1 = \omega_4 = 1.11$, $\omega_2 = \omega_3 = 1.15$, $\omega_c = 1.15$, $g_1 = g_4 = 0.55$, $g_2 = 0.52$, $g_3 = 0.5$, $\nu_1 = \nu_3 = 2.2$, $\nu_2 = \nu_4 = 2.4$ and $q = 0.75$. For the QDs, we have taken $\theta_i = \{\pi/4, 3\pi/5, 2\pi/3, 3\pi/4\}$ and $\phi_i = \{3\pi/4, \pi/3, \pi/4, \pi/6\}$.}
	\label{fig:tc_eterna_NM}
\end{figure}
In Fig. \ref{fig:tc_eterna_NM}, we have plotted the dynamics of the spin system through phase covariant eternal non-Markovian (PCEnM) master equation, Eq. (\ref{Eq:phase-cov-eternal}). It is observed that under the influence of the PCEnM channel, which is an eternal CP-indivisible non-Markovian channel for $t > 0$ (except when $t\to\infty$), the $P$ function takes negative values for longer periods of time. This indicates that quantumness in the system is retained for a longer time. Further, in Fig. \ref{fig:tc_eterna_NM}, variation of the $P$ function for a limiting value of the pure dephasing rate $\gamma_{3k}(t \to\infty) = -\nu_k/2$ is plotted. It can be observed that at shorter times, the $P$ functions for time-dependent $\gamma_{3k}(t)$ (black dot-dashed curve) and constant $\gamma_{3k}$ (green dashed curve) are slightly different. However, as the time-dependent $\gamma_{3k}(t)$ converges to $-\nu_k/2$, the $P$ functions coincide. The difference in the evolution of the $P$ function is due to the structure of the PCEnM channel, which is eternally non-Markovian when $\gamma_{3k}(t)$ is time-dependent and Markovian when it becomes time-independent in the limit $t\to\infty$.

\subsubsection{\label{sec:NMAD}Non-Markovian amplitude damping channel}
Here we consider the non-Markovian amplitude damping (NMAD) channel~\cite{nmad-channel, nmad-garraway} for the evolution of the spin system. To this end, the master equation for the evolution of the system is given by 
\begin{align}
\frac{d\rho(t)}{dt} &= -i[H, \rho] + \sum_{k=1}^N\gamma_k'(t)[\sigma_k^-\rho\sigma_k^+ - \frac{1}{2}\{\sigma_k^+\sigma_k^-, \rho(t)\}] \nonumber \\
&+ \frac{1}{2} \kappa (2a \rho(t) a^{\dagger} - a^{\dagger} a\rho(t) - \rho(t) a^{\dagger}a),
\label{Eq:NMAD-masterEq}
\end{align}
where $\gamma_k'(t) = -2\Re\left(\frac{\dot{F}(t)}{F(t)}\right)$ is the time dependent decoherence rate. $F$ is the decoherence function given by 
\begin{equation}
    F(t) = e^{\frac{-q't}{2}}\left(\frac{q'}{l}\sinh\left[\frac{lt}{2}\right] + \cosh\left[\frac{lt}{2}\right]\right),
\end{equation}
with $l = \sqrt{q^{'2} - 2\gamma'_k q'}$. Here $\gamma'_k$ and $q'$ parameterize the bath spectral density~\cite{nmad-channel}. Now
\begin{align}
    \gamma_k'(t)  &= -\frac{2}{|F(t)|}\frac{d|F(t)|}{dt} \nonumber \\
    &= 2\Re\left( \frac{\gamma'_k}{\sqrt{1 - \frac{2\gamma'_k}{q'}}\coth\left(\frac{1}{2}q't\sqrt{1 - \frac{2\gamma'_k}{q'}}\right) + 1}\right). 
\end{align}
\begin{figure}[h]
	\includegraphics[height=85mm,width=1\columnwidth]{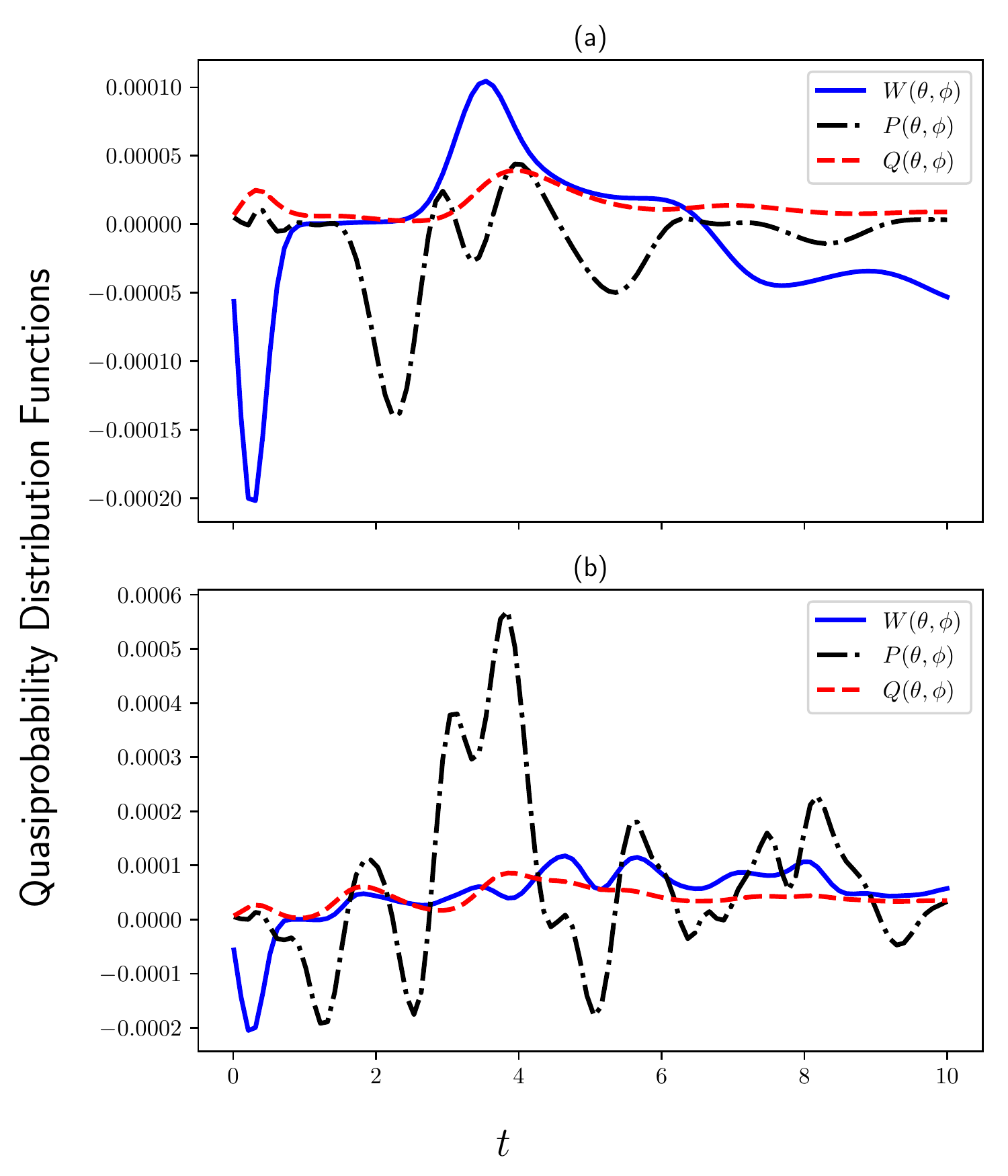}
	\caption{Variation of the $W$, $P$, and $Q$ function for the spin system with time. Subplot (a) shows the QDs for the spin system after the GKSL type of evolution, modeling an AD channel, and (b) shows the QDs for the spin system after evolution through the NMAD channel given in Eq. (\ref{Eq:NMAD-masterEq}). The parameters have the following values: $\omega_1 = \omega_4 = 1.11$, $\omega_2 = \omega_3 = 1.15$, $\omega_c = 1.15$, $g_1 = g_4 = 0.55$, $g_2 = 0.52$, $g_3 = 0.5$, $\gamma_i = 0.5 + g_i$, $\gamma'_i = \gamma_i/2$, $q' = 0.05$ and $\kappa  =0.01$. For the QDs, we have taken $\theta_i = \{\pi/4, 3\pi/5, 2\pi/3, 3\pi/4\}$ and $\phi_i = \{3\pi/4, \pi/3, \pi/4, \pi/6\}$.}
	\label{fig:tc_NMAD}
\end{figure}
The evolution, in this case, is non-Markovian for $q'<2\gamma'_k$ as that regime leads to damped oscillations. Here $\Re$ represents the real part of the quantity inside the bracket.
In Fig. \ref{fig:tc_NMAD}, we have plotted the evolution of the spin system through the GKSL master equation (modeling an AD channel), Fig. \ref{fig:tc_NMAD} (a), and through the NMAD channel, Fig. \ref{fig:tc_NMAD} (b). We observe that under the influence of the NMAD channel, the $P$ function becomes negative before the corresponding case under the GKSL master equation. 

\subsubsection{\label{sec:semi-Markov}Semi-Markov dephasing channel}
We now discuss the evolution of the spin system through the semi-Markov dephasing channel~\cite{semi-markov-channel} using the master equation
\begin{align}
    \frac{d \rho(t)}{dt} &= -i[H, \rho] +\sum_{k=1}^N {\tilde \gamma}_k (t)[\sigma_k^z\rho\sigma_k^z - \rho] \nonumber \\
    &+ \frac{1}{2} \kappa (2a \rho(t) a^{\dagger} - a^{\dagger} a\rho(t) - \rho(t) a^{\dagger}a),
    \label{Eq:semi-markov}
\end{align}
where 
\begin{align}
    {\tilde\gamma_k(t)} = \frac{2\tilde\gamma_k}{s\sqrt{1 - \frac{8\tilde\gamma_k}{s^2}}\coth\left(\frac{1}{2}st\sqrt{1 - \frac{8\tilde\gamma_k}{s^2}}\right) + s}. 
\end{align}
\begin{figure}[h]
	\includegraphics[height=85mm,width=1\columnwidth]{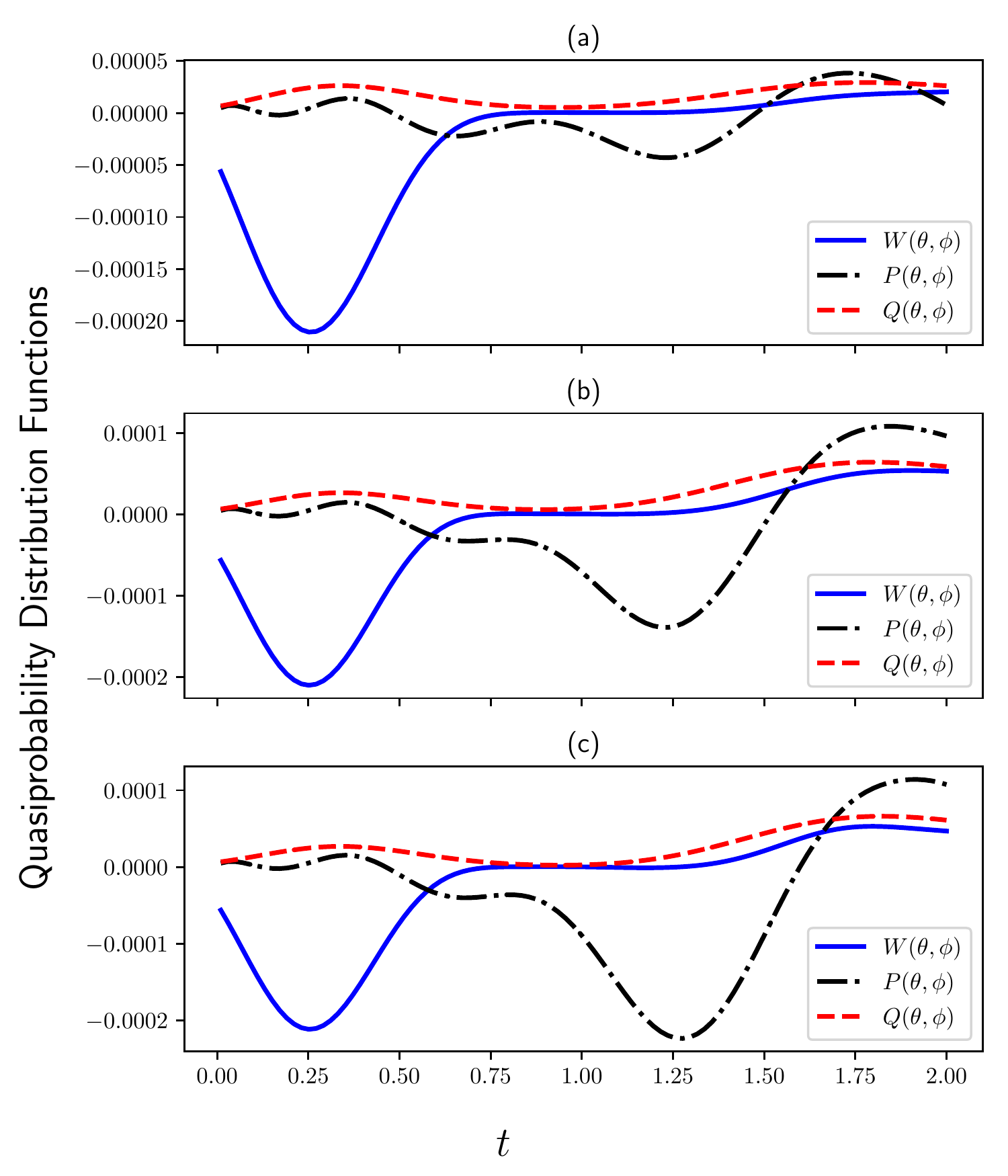}
	\caption{Variation of the $W$, $P$, and $Q$ function for the spin system with time. Subplot (a) shows the QDs for the spin system after Lindblad type evolution, modeling an AD channel, and (b) shows the QDs for the spin system after evolution through semi-Markov dephasing channel given in Eq. (\ref{Eq:semi-markov}). The parameters are chosen to be: $\omega_1 = \omega_4 = 1.11$, $\omega_2 = \omega_3 = 1.15$, $\omega_c = 1.15$, $g_1 = g_4 = 0.55$, $g_2 = 0.52$, $g_3 = 0.5$, $\gamma_1 = \gamma_3 = 0.31$, $\gamma_2 = \gamma_4 = 0.32$, $\tilde\gamma_i = \gamma_i/2$, $s=0.1$ and $\kappa = 0.01$. Also, for the QDs, we have taken $\theta_i = \{\pi/4, 3\pi/5, 2\pi/3, 3\pi/4\}$ and $\phi_i = \{3\pi/4, \pi/3, \pi/4, \pi/6\}$. Subplot (c) shows the variation of QDs for pure Hamiltonian evolution of the system with $\tilde \gamma_k = \kappa = 0$.}
	\label{fig:tc_semi_Markov}
\end{figure}
In this case, the evolution is CP-indivisible and non-Markovian for $p>\frac{s^2}{8}$.
From Fig. \ref{fig:tc_semi_Markov}, we observe that the evolution of the QDs for the spin system is nearly the same under the evolution of the system through the semi-Markov dephasing channel and through the GKSL master equation, modeling an AD channel. 
Further, we also plotted the variation of QDs when the system's evolution is through the pure Hamiltonian ($\tilde \gamma_k = \kappa = 0$) in Fig. \ref{fig:tc_semi_Markov}(c). It can be observed that the $P$ function is the most negative when the system is evolved under pure Hamiltonian, and the negativity of the $P$ function reduces slightly when evolution is under the semi-Markov dephasing channel. The $P$ function is the least negative when the AD channel is used for the system's evolution.

\begin{figure}[h]
	\includegraphics[height=75mm,width=1\columnwidth]{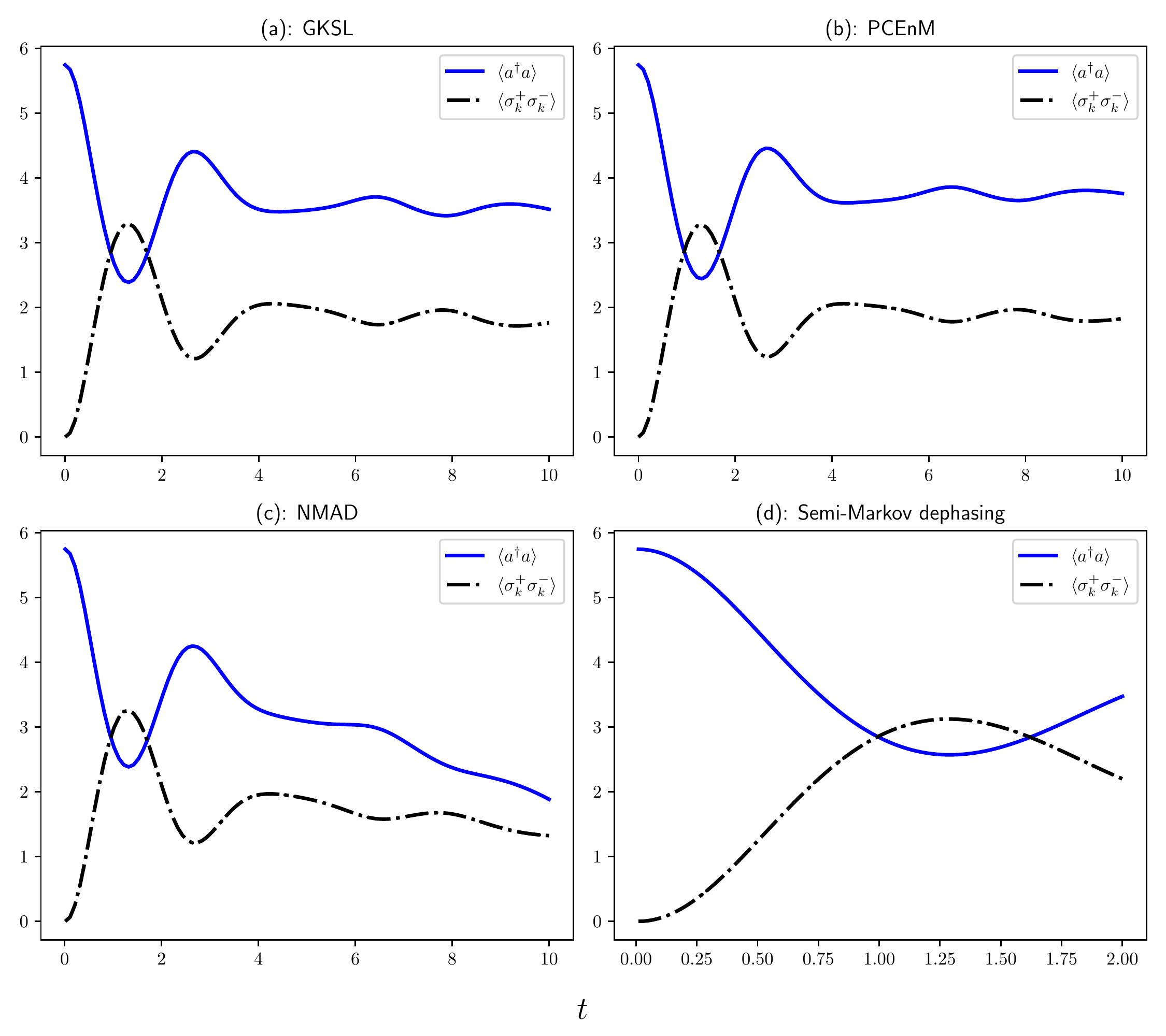}
	\caption{Variation of the cavity photon number $\langle a^{\dagger}a\rangle$ and the spin excitation $\langle\sigma^+_k\sigma^-_k\rangle$ with time where the evolution of the spin system is under different quantum channels: (a) through GKSL master equation (Eq. (\ref{Eq:GKSL-master-eq})) with parameters: $\gamma_i = 0.01g_i$, (b) under PCEnM channel (Eq. (\ref{Eq:phase-cov-eternal})) (with parameters: $\nu_i = 0.02g_i$), (c) through NMAD channel (Eq. (\ref{Eq:NMAD-masterEq})) with parameters: $\gamma'_i = 0.5 + g_i/2$, $q' = 0.05$, and (d) semi-Markov dephasing channel (Eq. (\ref{Eq:semi-markov})) with parameters: $\tilde\gamma_1 = \tilde\gamma_3 = 0.31/2$, $\tilde\gamma_2 = \tilde\gamma_4 = 0.32/2$, $s=0.1$. Values of $g_1 = g_4 = 0.55$, $g_2 = 0.52$, $g_3 = 0.5$, $\omega_1 = \omega_4 = 1.11$, $\omega_2 = \omega_3 = 1.15$, $\omega_c = 1.15$ and $\kappa=0.01$ are same for all the channels.}
	\label{fig:tc_number_spsm}
\end{figure}
We will next examine the dynamics of the cavity field. As a connection to the spin dynamics studied till now, we observe the effect of the evolution of the spin-cavity system due to the impact of various quantum channels using the spin excitation $\langle\sigma^+_k\sigma^-_k\rangle$ (where $k$ runs from 1 to $N$) and cavity photon number $\langle a^{\dagger}a\rangle$, and is depicted in Fig. \ref{fig:tc_number_spsm}. 
The peaks and dips in the spin excitation are in contrast with those in the cavity photon number, indicating a transfer of excitation from the spins to the cavity. This could be attributed to the fact that total excitation number $\sum_{k=1}^4 \sigma^+_k \sigma^-_k + a^\dagger a$ is conserved in the TC model (for negligible dissipation, as for the parameters chosen in the present case) and serves as a consistency check. This was also observed in the context of a mesoscopic spin-cavity system~\cite{hsdhar}. 
 
\section{\label{sec:cavity-dynamics}Dynamics of cavity field}
\begin{figure}[h]
	\includegraphics[height=60mm,width=1\columnwidth]{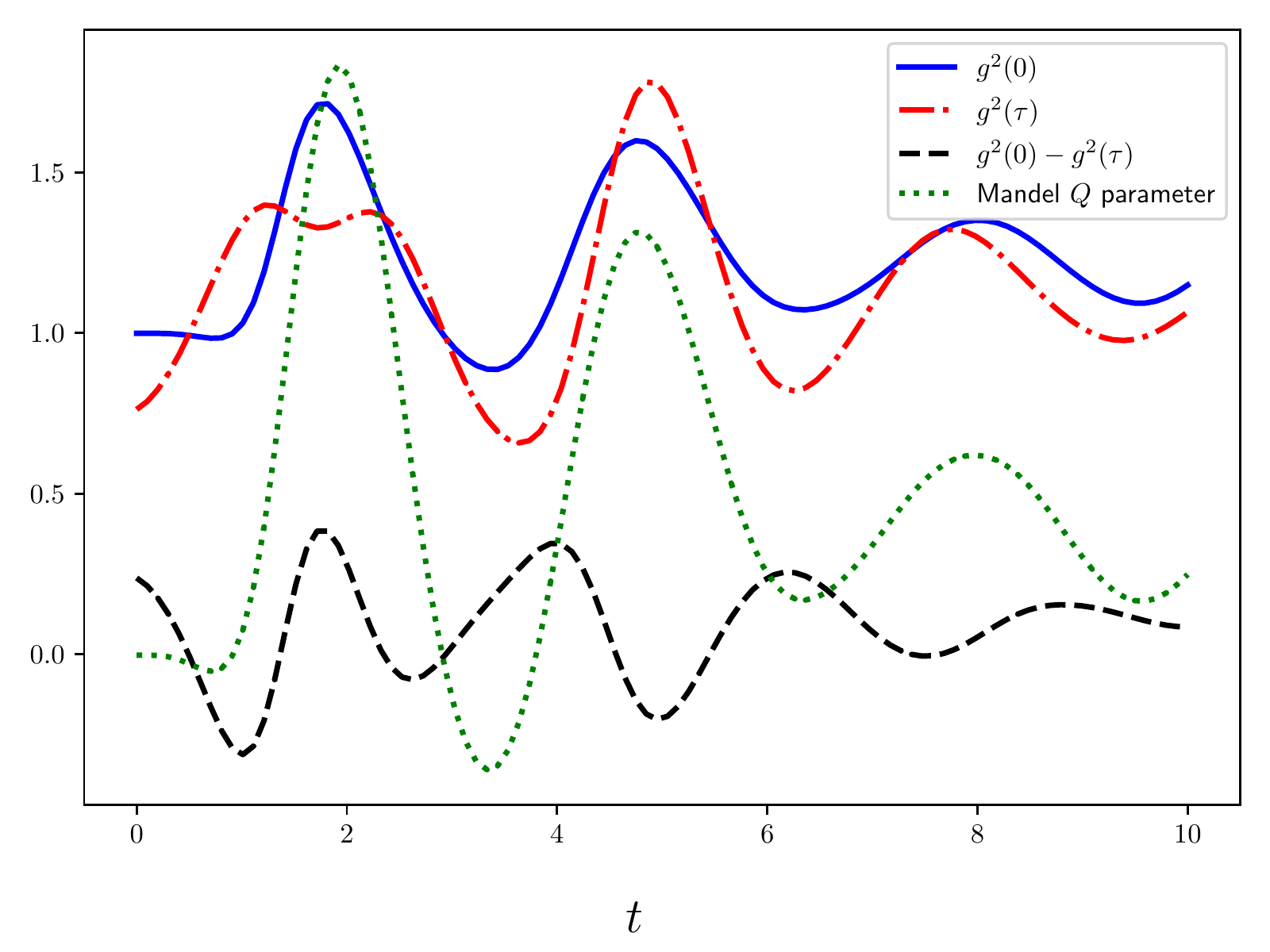}
	\caption{Variation of Mandel Q parameter and the second-order coherence function $g^{(2)}(\tau)$ (for $\tau=0$ and $\tau=3$) and the difference $g^{(2)}(0) - g^{(2)}(\tau)$  with time. In this case, the cavity's evolution is considered through the GKSL master equation. The parameters are: $\omega_1 = \omega_4 = 1.11$, $\omega_2 = \omega_3 = 1.15$, $\omega_c = 1.15$, $g_1 = g_4 = 0.55$, $g_2 = 0.52$, $g_3 = 0.5$, $\gamma_i = 0.01g_i$, $\kappa = 0.1$.}
	\label{fig:g2-tau}
\end{figure}

In this section, we study the dynamics of the cavity field. To this end, we use the second-order coherence function to characterize the evolution of the field. The second-order coherence function ($g^{(2)}$) is one of the most important characterizers of a light source into classical or non-classical and bunched or anti-bunched. It is defined as 
\begin{equation}
    g^{(2)}(\tau) = \frac{\langle \hat{a}^{\dagger}(t)\hat{a}^{\dagger}(t+\tau)\hat{a}(t+\tau)\hat{a}(t)\rangle}{\langle \hat{a}^{\dagger}(t)\hat{a}(t)\rangle^2},
\end{equation}
where $\hat{a}$ and $\hat{a}^{\dagger}$ are the bosonic annihilation and creation operators, respectively, representing the cavity field. 
Here we calculate the $g^{(2)}(0)$ function after the evolution of the cavity field through the GKSL master equation (AD channel) and make a comparison with the AD channel being replaced by the NMAD channel, using the following master equation
\begin{align}
    \frac{d \rho(t)}{dt}&= -i[H, \rho] + \sum_{k=1}^N \gamma_k (2\sigma_k^-\rho(t)\sigma_k^+ - \{\sigma_k^+\sigma_k^-, \rho(t)\}) \nonumber \\
    & + \kappa'(t)[\hat{a}\rho(t)\hat{a}^{\dagger} - \frac{1}{2}\{\hat{a}^{\dagger}\hat{a}, \rho(t)\}],
\end{align}%
where 
\begin{equation}
    \kappa'(t) = 2\Re \left( \frac{\kappa'}{\sqrt{1 - \frac{2\kappa'}{b}}\coth\left(\frac{1}{2}bt\sqrt{1 - \frac{2\kappa'}{b}}\right) + 1}\right).
\end{equation}
Here, we have chosen $b<2\kappa'$ for non-Markovian evolution, and $\Re$ denotes the real part of the quantity inside the bracket.
\begin{figure}[h]
	\includegraphics[height=60mm,width=1\columnwidth]{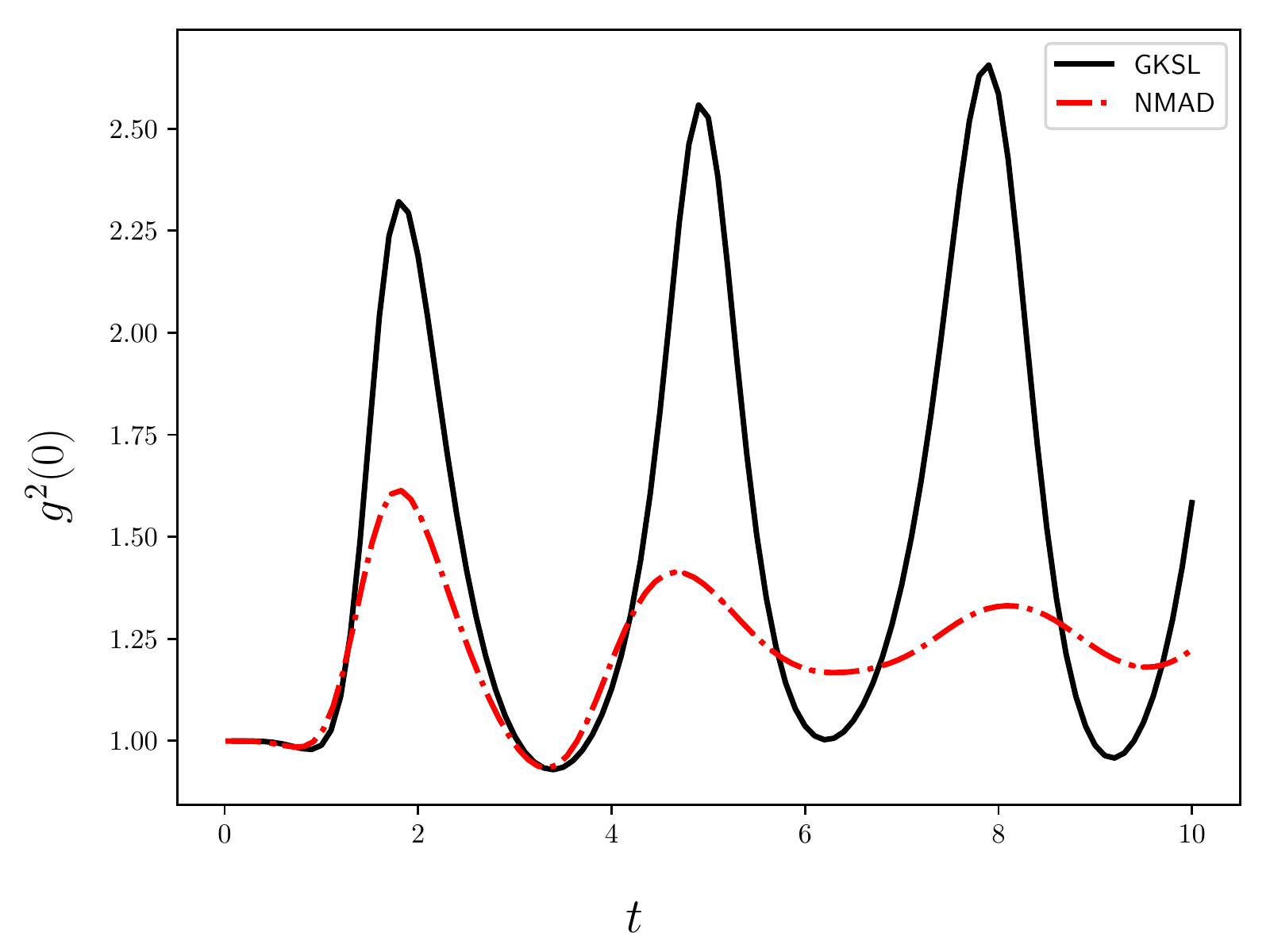}
	\caption{Variation of the equal-time second-order coherence function $g^{(2)}(0)$ with time. The black solid line refers to the evolution of the cavity mode through the GKSL master equation (AD channel), and the dot-dashed red line refers to the corresponding evolution under the NMAD channel. The parameters are: $\omega_1 = \omega_4 = 1.11$, $\omega_2 = \omega_3 = 1.15$, $\omega_c = 1.15$, $g_1 = g_4 = 0.55$, $g_2 = 0.52$, $g_3 = 0.5$, $\gamma_i = 0.1g_i$, $\kappa = 0.5$, $\kappa' = \kappa/2$, and $b=0.05$.}
	\label{fig:cavity-NMAD}
\end{figure}
In Fig. \ref{fig:cavity-NMAD}, we draw a comparison between the evolution of the cavity field impacted by the AD and NMAD channels. In both cases, we observe a similar pattern of dips and blips in the $g^{(2)}(0)$ function. The $g^{(2)}(0)$ is observed to be less than $1$ a number of times, indicating the sub-Poissonian (non-classical) nature of light. A decay in the $g^{(2)}(0)$ function sets in earlier for evolution under the NMAD channel as compared to that under the AD channel. 
An important parameter characterizing the light source is the Mandel Q parameter. It is defined as 
\begin{equation}
    \text{Q} = \langle\hat{n}\rangle(g^{(2)}(0) - 1),
\end{equation}
where $\hat{n}$ is the photon number operator. In Fig. \ref{fig:g2-tau}, the negative values of the Mandel Q parameter depict the sub-Poissonian behavior of light.

We finally calculate the $g^{(2)}(\tau)$ function, shown in Fig. \ref{fig:g2-tau}. Light is said to be anti-bunched if $g^{(2)}(0)<g^{(2)}(\tau)$ and bunched if $g^{(2)}(0)>g^{(2)}(\tau)$~\cite{loudon}. 
In Fig. \ref{fig:g2-tau}, we observe that the light is anti-bunched a number of times. Also, at around $t=5$, we observe that the light becomes anti-bunched ($g^{(2)}(0) - g^{(2)}(\tau)<0$) even when the $g^{(2)}(0)$ function is not less than 1. This brings out the difference between the anti-bunched and sub-Poissonian behavior of light~\cite{zou-mandel} in the present context.

\section{\label{sec:conclusion}Conclusion}
We have discussed the Tavis-Cummings model in a noisy environment. The impact of both Markovian and non-Markovian noise on the dynamics of the Tavis-Cummings model was observed. The dynamics of the spin system and the cavity field were studied using the $W$, $P$, and $Q$ quasi-probability distributions and second-order coherence function ($g^{(2)}$ function), respectively. 
The effect of squeezing and temperature on the quasi-probability distributions was analyzed.
We observed that, in general, an increase in the squeezing parameters and temperature diminishes non-classicality in the system for a given set of atomic parameters. We have studied the impact of PCEnM, NMAD, and semi-Markov dephasing channels, in their non-Markovian limits, over the evolution of the atomic system using the $W$, $P$, and $Q$ functions. We have compared the impact of NMAD and semi-Markov dephasing channels with the variation of these quasi-probability distribution functions when the atomic system is evolved through the GKSL master equation modeling an AD channel, which is semi-group and models a Markovian evolution. This brings out the impact of  memory effects on the quasiprobability distribution functions.
From the study of the influence of noise on the spin-cavity system, specifically on the spin excitation and the cavity photon number, it was observed that, in general, the spin excitation and the cavity photon number have a complementary behavior.
The dynamics of the Mandel Q parameter revealed the cavity field to be sub-Poissonian for different evolution times for the parameters considered. 
The $g^{(2)}(\tau)$ function was also computed and compared with the $g^{(2)}(0)$ function in order to bring out bunching and anti-bunching in the light. Interestingly, a number of instances were observed where sub-Poissonian and anti-bunching behavior of light were not in tandem.
\section*{Acknowledgements}
The authors acknowledge useful discussions with Himadri Shekhar Dhar during the preliminary stage of the work. SB acknowledges support from the Interdisciplinary Cyber Physical Systems (ICPS) programme of the Department of Science and Technology (DST),
India, Grant No.: DST/ICPS/QuST/Theme-1/2019/6. SB also acknowledges support from the Interdisciplinary Research Platform (IDRP) on Quantum Information and Computation (QIC) at IIT Jodhpur.

\bibliographystyle{apsrev4-1}
\bibliography{reference}

\end{document}